\documentclass[a4paper,11pt,twoside]{article}
\usepackage{latexsym}
\usepackage{amsmath}
\usepackage{amsfonts}
\usepackage{amssymb}
\usepackage{vmargin}
\usepackage{graphicx}
\usepackage{palatino}
\usepackage{url}
\usepackage[all]{nowidow}
\usepackage{fancyhdr}
\usepackage{subcaption}
\usepackage[font=small]{caption}

\newcommand*{\Scale}[2][4]{\scalebox{#1}{$#2$}}%
\numberwithin{equation}{section}

\setmarginsrb{2.7cm}{3cm}{2.7cm}{4cm}{0mm}{10mm}{0mm}{0mm}
\makeatletter
\renewcommand{\maketitle}{\bgroup\setlength{\parindent}{0pt}
\begin{flushleft}
{\footnotesize 
\vspace*{-2cm}
Hindawi Publishing Corporation\\
Journal of Applied Mathematics\\
Volume 2011, Article ID 863161, 11 pages\\
doi: 10.1155/2011/863161 \par}  \vspace*{1 cm}

{\LARGE
\textit{\Large Research Article}\\

\textbf{\@title} \par}
\vspace{1cm}
\@author
\end{flushleft}\egroup
}
\makeatother

\title{\bfseries Stability of the NLS Equation with Viscosity Effect}
\date{\small }
\author{\hspace*{1cm} \normalsize \textbf{N. Karjanto$^\mathbf{\scriptsize 1}$ and K. M. Tiong$^\mathbf{\scriptsize 2}$}}

\begin{document}
\maketitle
\setlength{\leftskip}{1cm}
\noindent 
\textit{\small $^{\scriptsize 1}$Department of Mathematics, University College, Sungkyunkwan University\\ 
\hspace*{1mm} Natural Science Campus, Jangan-gu, Suwon, 16419, Gyeonggi-do, Republic of Korea}\\
\textit{\small $^{\scriptsize 2}$Nottingham University Business School, The University of Nottingham Malaysia Campus\\ 
\hspace*{1mm} Jalan Broga, Semenyih, 43500, Selangor, Malaysia} \vspace{0.25cm}\\
{\small Correspondence should be addressed to N. Karjanto, \ {\footnotesize \url{natanael@skku.edu}}} \vspace{0.25cm}\\
{\small Received 16 November 2010; Accepted 11 January 2011} \vspace{0.25cm}\\
{\small Academic Editor: Ramon Codina}  \vspace{0.25cm}\\
{\small Copyright \copyright \ 2011 $\;$ N. Karjanto and K. M. Tiong. This is an open access article distributed under the Creative Commons Attribution License, which permits unrestricted use, distribution, and reproduction in any medium, provided the original works is properly cited. \par}  \vspace{0.5cm}
\noindent
{\small 
A nonlinear Schr\"{o}dinger (NLS) equation with an effect of viscosity is derived from a Korteweg-de Vries (KdV) equation modified with viscosity using the method of multiple time-scales. It is well known that the plane-wave solution of the NLS equation exhibits modulational instability phenomenon. In this paper, the modulational instability of the plane-wave solution of the NLS equation modified with viscosity is investigated. The corresponding modulational dispersion relation is expressed as a quadratic equation with complex-valued coefficients. By restricting the modulational wavenumber into the case of narrow-banded spectra, it is observed that a type of dissipation, in this case, the effect of viscosity, stabilizes the modulational instability, as confirmed by earlier findings.
\par} \vspace*{0.5cm}

\setlength{\leftskip}{0pt}
\setlength{\abovedisplayshortskip}{15pt}
\setlength{\belowdisplayshortskip}{15pt} 
\setlength{\abovedisplayskip}{15pt}
\setlength{\belowdisplayskip}{15pt}
\section{Introduction}

Modulational instability is a well-known phenomenon in the context of nonlinear science, particularly those systems that have solutions in waveforms. In the context of water waves, it is the growth in space or in time of a wave amplitude under a small wave modulation~\cite{Scott05}. It is also known as sideband instability or Benjamin-Feir instability since they investigated the phenomenon of deep-water wave trains~\cite{BenjFeir67}. In the field of nonlinear optics, modulational instability is a phenomenon whereby deviations from an optical waveform are reinforced by nonlinearity, leading to the generation of spectral-sidebands and the eventual breakup of the waveform into a train of pulses~\cite{Agrawal95}.

The nonlinear Schr\"{o}dinger (NLS) equation provides a canonical description for the modulational instability of a weakly nonlinear quasi-monochromatic dispersive wave packet. In particular, we are interested in the focusing type of the NLS equation. In this paper, we would like to investigate the stability of the plane-wave solution of the NLS equation where an effect of viscosity is present. This modified NLS equation includes both the dissipative and dispersive effects of viscous boundary layers. It describes the evolution equation for the envelope of weakly nonlinear long waves on the free surface with a finite depth. The NLS equation with viscosity effect can be derived using multiple space- and time-scales method from the KdV equation that has a similar effect. While the previous equation is not really well known in the literature, the latter one has been discussed and investigated by a number of authors.

Back in the late 1960s, a theory describing the oscillations of a liquid tank near resonant frequency is developed by Chester~\cite{Chester68}. The author featured the significant effects of dissipation and dispersion to determine oscillations, with the formation of a weak bore which travels to and fro in the tank when the former case is dominant and a series of traveling cnoidal waves along the tank when the latter case is significant. Later, a modified KdV equation to include energy dissipation is derived in the 1970s by Ott and Sudan~\cite{OttSudan70}. The authors consider four different cases of the presence of dissipation: magnetosonic waves damped by electron-ion collision, ion-sound waves damped by ion-neutral collisions, ion-sound waves with electron Landau damping and shallow water waves damped by viscosity. The effect of a viscous boundary layer along the sidewall and the bottom on the propagation of infinitesimal surface gravity waves is also studied in the 1970s by Mei and Liu~\cite{MeiLiu73}.

Based on the governing equation developed by Chester~\cite{Chester68}, a KdV equation modified by viscosity is derived by Miles in the mid-1970s~\cite{Miles76}. This modified KdV equation has an extra term that describes the dissipative and dispersive effects of viscous boundary layers added to the classical KdV equation derived by Korteweg and de Vries at the end of the $19^\textmd{th}$ century~\cite{KdV1895}. Another derivation of a simple equation for weakly nonlinear long gravity waves on a viscous fluid layer is also obtained independently around the same period by Kakutani and Matsuuchi~\cite{KakuMats75}. Depending on the comparisons of wavelength and the Reynolds number, they observed that the viscosity effect does not affect the classical inviscid type of the KdV equation for ${\cal O}(k_{0}^{-5}) < \textmd{Re}$ while it modifies by adding an extra term for ${\cal O}(k_{0}^{-1}) < \textmd{Re} \leq {\cal O}(k_{0}^{-5})$, where $k_{0}$ is the wavenumber and $\textmd{Re}$ is the Reynolds number. The corresponding initial value problems for this type of KdV equation with viscosity effect are investigated numerically by Matsuuchi~\cite{Matsu76} and the obtained results are compared with the earlier experiments by Zabusky and Galvin~\cite{ZabuGal71}. Another interesting remark in term of water waves modeling, when a small viscosity is taken into account, the dissipative term is merely proportional to the wave amplitude~\cite{Lakeetal77,LakeYuen77}.

Furthermore, a KdV equation modified by viscosity effect to include internal-wave systems is derived by Koop and Butler in the early 1980s~\cite{KoopButl81}. Later, a KdV equation modified by viscosity for weakly nonlinear long waves propagating in a channel of uniform but arbitrary cross section using the method of matched asymptotic expansion is derived by Das~\cite{Das85} and the one for waves in a two-layer fluid by Das and Chakrabarti~\cite{DasCha86}, both in the mid-1980s.

This paper is organized as follows. In Section~\ref{derivation}, the corresponding NLS equation with viscosity effect is derived from the KdV equation with viscosity effect from Miles~\cite{Miles76} using the method of multiple time-scales. In Section~\ref{stability}, we investigate the stability of the plane-wave solution of the derived NLS equation with viscosity effect, providing the growth rate as a function of wavenumber and its modulation. Finally, Section~\ref{conclusion} gives a conclusion to our discussion.

\section{Derivation of the NLS equation with Viscosity Effect} \label{derivation}
Consider the KdV equation with viscosity effect given by Miles~\cite{Miles76},
\begin{equation}
  f_{\tau} + f f_{\xi} + \epsilon \bar{\alpha} f_{\xi \xi \xi} + \bar{\beta} \partial_{\xi}(\textmd{$\cal{D}$} f) = 0, \label{KdVMiles}
\end{equation}
where $f$ is the surface wave elevation, $\bar{\alpha}$ and $\bar{\beta}$ are both of $\cal{O}$(1), and $\cal{D}$ is a diffusion operator and is given by
\begin{equation}
  \textmd{$\cal{D}$} f = \int_{0}^{\infty} \frac{1}{\sqrt{\pi \zeta}} f(\xi - \zeta, \tau)\, d\zeta.
\end{equation}
By writing $\eta(x,t) = \epsilon f(\xi,\tau)$, $\xi = \epsilon(x - ct)$ and $\tau = \epsilon^{2} t$ as the moving of reference, then we have $\partial_{x} = \epsilon \partial_{\xi}$ and $\partial_{t} = - c \epsilon \partial_{\xi} + \epsilon^{2} \partial{\tau}$. Thus, $f_{\xi} = (1/\epsilon^{2}) \eta_{x}$, $f_{\tau} = (1/\epsilon^{3})(\eta_{t} + c \eta_{x})$, $f_{\xi \xi \xi} = (1/\epsilon^{4}) \eta_{xxx}$ and, by taking $s = \epsilon^{2} \zeta$, the last term of~\eqref{KdVMiles} becomes
\begin{equation}
  \partial_{\xi}(\textmd{$\cal{D}$} f) = \frac{1}{\epsilon^{3}} \frac{\partial}{\partial x} \int_{0}^{\infty} \frac{1}{\sqrt{\pi s}} \, \eta \left(x - \frac{s}{\epsilon^{3}},t \right) \, ds.
\end{equation}
The corresponding equation at the order of $\epsilon^{-3}$ is given as follows:
\begin{equation}
  \eta_{t} + c \eta_{x} + \eta \eta_{x} + \bar{\alpha} \, \eta_{xxx} + \bar{\beta} \, \partial_{x}(\textmd{$\cal{D}$} \eta) = 0,
  \label{KdVvis}
\end{equation}
where the diffusion operator $\cal{D}$ is now given by
\begin{equation}
  \textmd{$\cal{D}$} \eta = \int_{0}^{\infty} \frac{1}{\sqrt{\pi s}} \eta \left(x - \frac{s}{\epsilon^{3}}, t \right)\, ds.
\end{equation}

Now, substitute the small surface elevation $\eta(x,t)$ as a wave packet expression consisting of a superposition of a first-order harmonic, a second-order double harmonic and a second-order non-harmonic long wave, given as follows:
\begin{equation}
  \eta(x,t) = \epsilon A(\xi,\tau) e^{i\theta_{0}} + \epsilon^{2} [B(\xi,\tau) e^{2i\theta_{0}} + C(\xi,\tau)] + \textmd{complex conjugate,}
\end{equation}
where $\theta_{0} = k_{0}x - \omega_{0}t$, $\xi = \epsilon(x - ct)$ and $\tau = \epsilon^{2}t$ are the variables in a moving frame of reference. The complex-valued amplitudes $A$, $B$ and $C$ are allowed to vary slowly in the frame of reference. In this case, $s = {\cal O}(\epsilon^{3})$ and thus $s/\epsilon^{3} = {\cal O}(1)$. We want to apply the method of multiple scales to find an evolution equation for the complex-valued amplitude of the lowest order harmonic $A(\xi,\tau)$. Substituting $\eta$ into~\eqref{KdVvis}, collecting the terms according to the order of $\epsilon$ and the exponential term $\theta_{0}$, we obtain the corresponding nonlinear Schr\"{o}dinger equation with an effect of viscosity at the lowest order $\cal{O}$(1) and the coefficient of the term $e^{i\theta_{0}}$, given as follows:
\begin{equation}
  \partial_{\tau} A + i \beta \partial_{\xi}^{2} A + i \gamma |A|^{2} A + \alpha \partial_{\xi} (\textmd{$\cal{D}$} A) = 0, \label{NLSvis0}
\end{equation}
where now the diffusion operator is given as follows:
\begin{equation}
  \textmd{$\cal{D}$} A = \int_{0}^{\infty} \frac{1}{\sqrt{\pi \zeta}} \, A(\xi - \zeta, \tau) e^{-i k_{0} \zeta/\epsilon}\, d\zeta.
\end{equation}
In this case, $\zeta = {\cal O}(\epsilon)$, and thus $\zeta/\epsilon = {\cal O}(1)$. The corresponding nonlinear dispersion relation is given by:
\begin{equation*}
  \omega = -\beta k^{2} + \gamma |A|^{2} + \alpha k (1 - i) \sqrt{\frac{\epsilon}{2(k_{0} + \epsilon k)}}.
\end{equation*}

This type of the NLS equation with viscosity effect seems to be novel and has never been investigated in the literature earlier. There are quite a number of published works which discuss wave propagation in a viscous fluid using the NLS equation with a dissipative term as its model. See for instance the works of Demiray~\cite{Demi01} and Tay {\slshape et al.}~\cite{Tay10} and the references therein, which describe models for wave propagation in an elastic tube with viscous fluid. For another type of dissipative NLS equation and its evolution of weakly nonlinear waves in the presence of viscosity and surfactant, see~\cite{Joo91}. Recently, a dissipative variant of NLS equation motivated by interactions of a thermal cloud with a Bose-Einstein condensates, as prototypically modeled by a damping term of the form described in~\cite{Cock10} is used for the dynamics of dark matter-wave solitons in elongated atomic condensates at finite temperatures. However, none of these models resembles the one described by~\eqref{NLSvis0}. In the following section, we will investigate the stability of the plane-wave solution of the NLS equation with viscosity effect.

\section{Modulational Instability} \label{stability}

There is a long list of references discussing on modulational instability related to the NLS equation and its variants. A simple and direct method to construct exact quasiperiodic solutions of the NLS equation and a study of the corresponding nonlinear modulational instability is discussed by Tracy {\slshape et al.} in the mid of 1980s~\cite{Tracy84}. Exact analytical expressions for modulational instability in the NLS equation are obtained by Akhmediev {\slshape et al.} around the same period~\cite{Akhmediev85,Akhmediev86}. The study of the modulational instability of short pulses propagating through long optical fibers by incorporating the effect of time derivative nonlinearity is presented by Shukla and Rasmussen~\cite{Shukla86}. A study of the modulation instability of an extended NLS is done by Potasek~\cite{Potasek87}. The author finds that although the time derivative of the nonlinearity affects the instability, it is discovered that  the odd-order higher dispersion does not contribute to the modulation instability frequency.

The possible effect of dissipation due to the viscosity effect of the water is considered by Tanaka~\cite{Tanaka90}, where the author compared the experimental result of Su and Green~\cite{SuGreen85} with the theoretical prediction given by the NLS equation. The author conjectured that the disagreement between theory and experiment for smaller values of the initial steepness might be contributed by the effect of viscosity. Another study of modulational instability in nonlocal nonlinear Kerr media is given by Krolikowski {\slshape et al.}~\cite{Krol01}. The authors show that for a focusing nonlinearity, the nonlocality can never remove the modulational instability completely irrespective of the particular profile of the nonlocal response function even though it tends to suppress it. The modulational instability in a cylindrically shaped Bose-Einstein condensates subject to one-dimensional optical lattices is shown both analytically and numerically by Konotop and Salermo~\cite{Kono02}.

The modulational instability of spatially uniform states in the NLS equation with the presence of higher-order dissipation term is discussed by Rapti {\slshape et al.}~\cite{Rapti04}. The authors show how the presence of even the weakest possible dissipation suppresses the instability on a longer time scale and confirm the result from Segur {\slshape et al.} who show that for waves with narrow bandwidth and moderate amplitude, any amount of dissipation (of a certain type) stabilizes the instability~\cite{Segur05}. On the other hand, however, Bridges and Dias pointed out that there is an overlooked mechanism whereby some kind of dissipation can enhance the modulational instability~\cite{BriDias07}. It is pointed out that due to the collision of modes with opposite energy sign, the negative energy perturbations are enhanced.

Another study on modulational instability for the NLS equation with a periodic potential is presented by Bronski and Rapti~\cite{BroRap05}. They discovered  that the small amplitude solutions corresponding to the band edges alternate stability, with the first band edge being modulationally unstable in the focusing case, the second band edge being modulationally unstable in the defocusing case, and so forth. The modulational instability of the higher-order NLS equation with fourth-order dispersion and quintic nonlinear terms describing the propagation of extremely short pulses is investigated by Hong~\cite{Hong06}.

Consider again the NLS equation with viscosity effect~\eqref{NLSvis0}. It can be checked that this NLS equation admits the plane-wave solution $A_{0}(\tau) = r_{0} e^{-i\gamma r_{0}^{2} \tau}$. This is due to the vanishing of the dispersive term and the diffusive term becomes
\begin{equation}
\Scale[0.97]{\displaystyle
{\cal D} A_{0} = A_{0} \int_{0}^{\infty} \frac{e^{-ik_{0} \zeta/\epsilon}}{\sqrt{\pi \zeta}} \, d\zeta = \frac{2}{\sqrt{\pi}} A_{0} \int_{0}^{\infty} e^{-ik_{0} u^{2}/\epsilon} \, du = \sqrt{\frac{\epsilon}{2 k_{0}}} (1 - i) A_{0} = \sqrt{\frac{\epsilon}{k_{0}}} A_{0} e^{-i\pi/4}.
}
\end{equation}
Taking derivative with respect to $\xi$ of this term also yields the vanishing term.

Substitute $A(\xi,\tau) = A_{0}(\tau)[1 + B(\xi,\tau)]$ into~\eqref{NLSvis0}, we obtain
\begin{align}
(\partial_{\tau}A_{0}) (1 + B) + A_{0}(\tau) \partial_{\tau} B + i \beta A_{0} \partial_{\xi}^{2} B + i \gamma |A_{0}|^{2} A_{0}(\tau) (1 + B) (1 + B + B^{\ast} + |B|^{2} ) \nonumber \\ + \; \alpha A_{0}(\tau) \partial_{\xi} ({\cal D}(1 + B)) = 0. \qquad \qquad \qquad \; \;
\end{align}
Simplifying the expression and ignoring the higher order terms yield a linearized NLS equation with viscosity effect
\begin{equation}
  \partial_{\tau} B + i \beta \partial_{\xi}^{2} B + i \gamma r_{0}^{2} (B + B^{\ast}) + \alpha \partial_{\xi} ({\cal D} B) = 0,  \label{linNLSvis}
\end{equation}
where $B^{\ast}$ is the complex conjugate of $B$ and
\begin{equation}
  {\cal D} B = \int_{0}^{\infty} \frac{1}{\sqrt{\pi \zeta}} \, B(\xi - \zeta, \tau) e^{-i k_{0} \zeta/\epsilon}\, d\zeta.
\end{equation}
Substitute an Ansatz $B(\xi, \tau) = B_{1} e^{(\omega \tau + i \kappa \xi)} + B_{2} e^{(\omega^{\ast} \tau - i \kappa \xi)}$ into the linearized NLS equation with viscosity effect~\eqref{linNLSvis}. In this case, $\omega \in \mathbb{C}$ is a growth rate and the coefficients $B_{1}$, $B_{2} \in \mathbb{C}$. The term ${\cal D} B$ now reads as
\begin{equation}
\begin{aligned}
{\cal D} B &= B_{1} e^{(\omega \tau + i \kappa \xi)} \int_{0}^{\infty} \frac{e^{-i(k_{0}/\epsilon + \kappa)\zeta}}{\sqrt{\pi \zeta}} \, d\zeta + 
			  B_{2} e^{(\omega^{\ast} \tau - i \kappa \xi)} \int_{0}^{\infty} \frac{e^{-i(k_{0}/\epsilon - \kappa)\zeta}}{\sqrt{\pi \zeta}} \, d\zeta \\
		   &= B_{1} T_{1}(\kappa) \sqrt{2} e^{-i\pi/4} e^{(\omega \tau + i \kappa \xi)} + 
			  B_{2} T_{2}(\kappa) \sqrt{2} e^{-i\pi/4} e^{(\omega^{\ast} \tau - i \kappa \xi)}
\end{aligned}
\end{equation}
where
\begin{equation}
  T_{1}(\kappa) = \frac{1}{\sqrt{2(k_{0}/\epsilon + \kappa)}} \qquad \qquad \textmd{and} \qquad \qquad
  T_{2}(\kappa) = \frac{1}{\sqrt{2(k_{0}/\epsilon - \kappa)}}.
\end{equation}
In particular, we are interested to consider the case where $0< \epsilon \kappa < k_{0}$ so that $T_{2}$ is defined and its value is always positive.

Then we have the following pair of coupled equations:
\begin{equation}
\begin{aligned}
\left[\omega - i \left(\beta \kappa^{2} - \gamma r_{0}^{2} \right) + \alpha \kappa T_{1}(\kappa) (1 + i) \right] B_{1} + i \gamma r_{0}^{2} B_{2}^{\ast} &= 0 \\
i \gamma r_{0}^{2} B_{1}^{\ast} + \left[\omega^{\ast} - i \beta \kappa^{2} + i \gamma r_{0}^{2} - \alpha \kappa T_{2} (\kappa)(1 + i) \right] B_{2}      &= 0.
\end{aligned}
\end{equation}
Taking complex conjugate of the second equation, we can write in matrix form
\begin{equation}
\Scale[0.9]{
\left[
  \begin{array}{cc}
    \omega - i \left(\beta \kappa^{2} - \gamma r_{0}^{2} \right) + \alpha \kappa T_{1} (\kappa)(1 + i) & i \gamma r_{0}^{2} \\
    -i \gamma r_{0}^{2} & \omega  + i \left(\beta \kappa^{2} - \gamma r_{0}^{2} \right) - \alpha \kappa T_{2} (\kappa)(1 - i)  \\
  \end{array}
\right]
\left[
  \begin{array}{c}
    B_{1} \\
    B_{2}^{\ast} \\
  \end{array}
\right]
=
\left[
  \begin{array}{c}
    0 \\
    0 \\
  \end{array}
\right].}
\end{equation}
The pair of linear homogeneous equations for $B_{1}$ and $B_{2}^{\ast}$ admits a nontrivial eigenvalue for $\omega$, provided that the determinant of the matrix coefficient is zero. We obtain the corresponding modulational dispersion relation, expressed as a quadratic equation in $\omega$ with complex-valued coefficients, given as follows:
\begin{equation}
  \omega^{2} + P_{1}(\kappa) \omega + P_{0}(\kappa) = 0, \label{quad}
\end{equation}
where the coefficients are complex-valued functions $P_{0}(\kappa)$ and $P_{1}(\kappa)$, given as follows:
\begin{equation}
\begin{aligned}
P_{0}(\kappa) &= \beta \kappa^{2} (\beta \kappa^{2} - 2 \gamma r_{0}^{2}) + (\beta \kappa^{2} - \gamma r_{0}^{2}) P_{1}(\kappa) 
				- 2 \alpha^{2} \kappa^{2} T_{1}(\kappa) T_{2}(\kappa) \\
P_{1}(\kappa) &= \alpha \kappa (T_{1}(\kappa) - T_{2}(\kappa) + i [T_{1}(\kappa) + T_{2}(\kappa)]).
\end{aligned}
\end{equation}
\begin{figure}[h]
\begin{center}
\begin{subfigure}{0.47\textwidth}
\includegraphics[width=\linewidth]{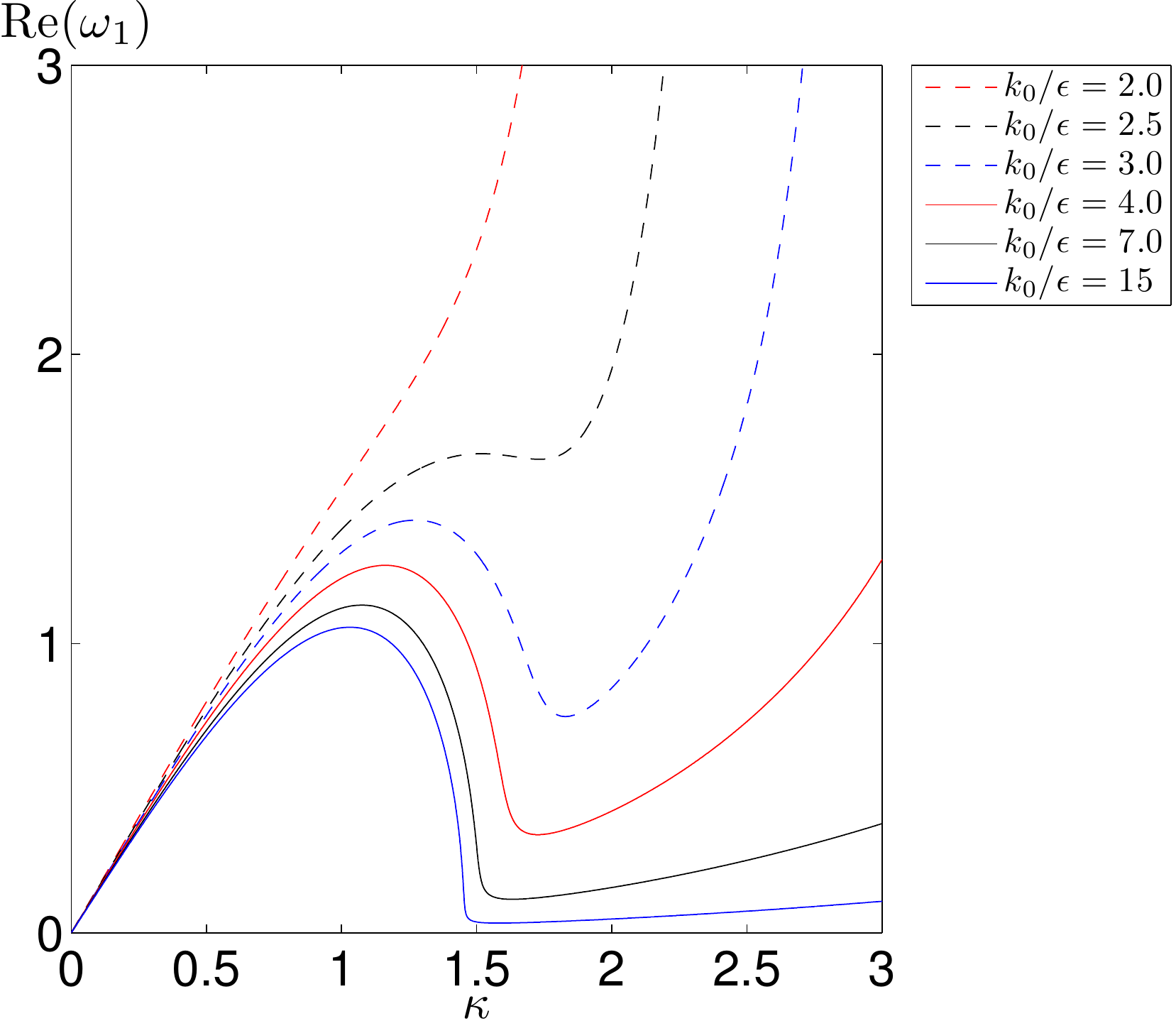} 
\caption{}
\end{subfigure}
\hspace{0.5cm}
\begin{subfigure}{0.47\textwidth}
\includegraphics[width=\linewidth]{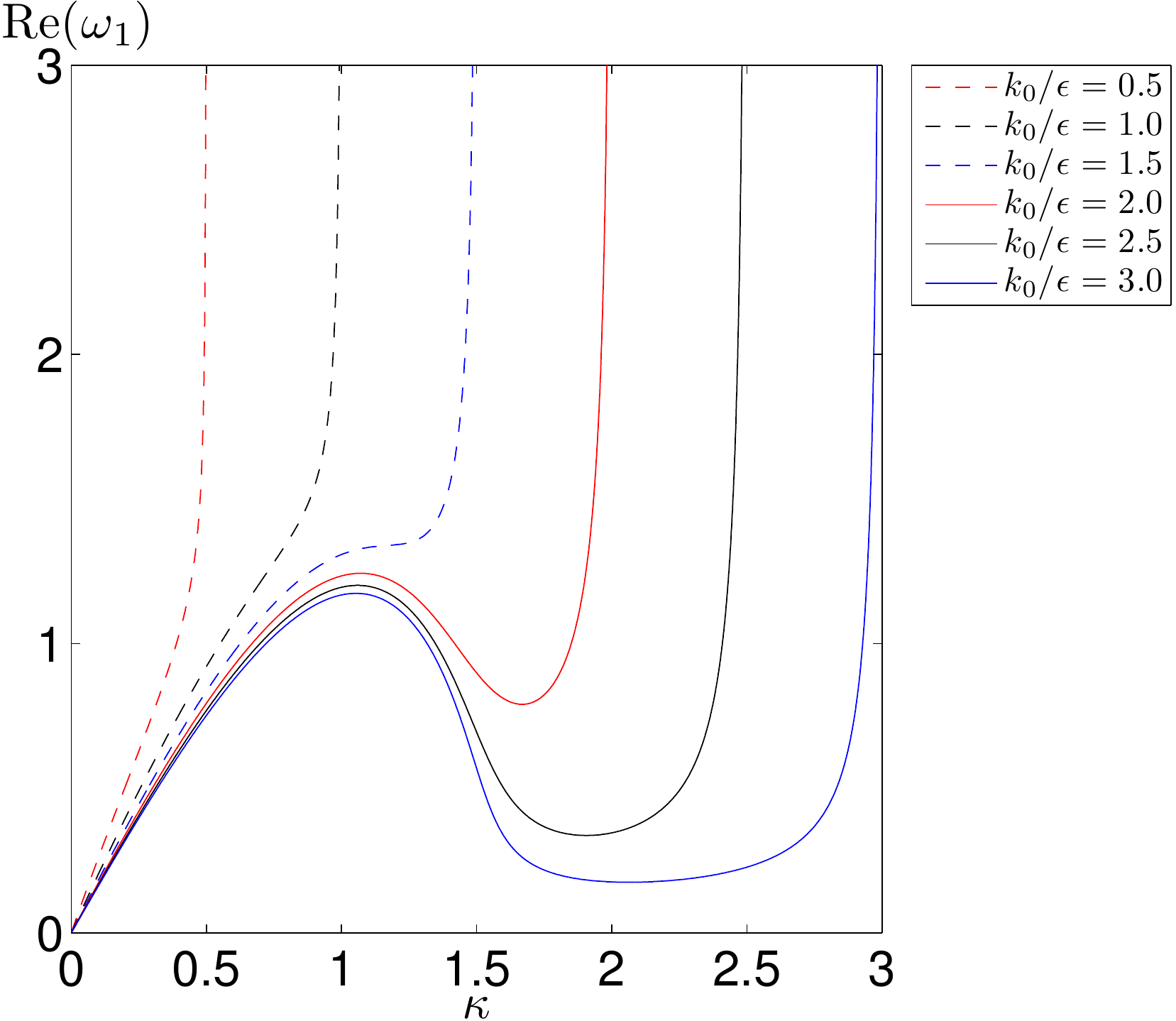}
\caption{}
\end{subfigure}
\caption{Plots of the real part of the growth rate $\omega_{1}$ as a function of the modulational wavenumber $\kappa$  for $\alpha = 1$ (a) and $\alpha = -1$ (b). The curves are given for several values of $k_{0}/\epsilon$.}
\label{pos}
\end{center}
\end{figure}

By completing the square of the modulational dispersion relation~\eqref{quad}, we obtain the following:
\begin{equation}
\left[2 \omega + P_{1}(\kappa) \right]^{2} = P_{1}^{2}(\kappa) - 4 P_{0}(\kappa).
\end{equation}
Furthermore, by writing the right-hand side of this expression in the polar form $R(\kappa) e^{i\theta(\kappa)}$, we obtain the solutions to~\eqref{quad} are given as follows:
\begin{equation}
\begin{aligned}
\omega_{1}(\kappa) &= \frac{1}{2} \left(\sqrt{R(\kappa)} e^{i \frac{1}{2} \theta(\kappa)} - P_{1}(\kappa)  \right), \\
\omega_{2}(\kappa) &= \frac{1}{2} \left(\sqrt{R(\kappa)} e^{i \left(\frac{1}{2} \theta(\kappa) + \pi\right)} - P_{1}(\kappa)  \right),
\end{aligned}
\end{equation}
where $R(\kappa)$ and $\theta(\kappa)$ satisfy the following expressions:
\begin{equation}
\begin{aligned}
\Scale[0.92]{R^{2}(\kappa) }	  &= \Scale[0.92]{[\textmd{Re}(P_{1}^{2} - 4P_{0})]^{2} + [\textmd{Im}(P_{1}^{2} - 4P_{0})]^{2}} \\
\Scale[0.92]{\tan \theta(\kappa)} &= 
\Scale[0.92]{\displaystyle \frac{2 \alpha^{2} \kappa^{2} (T_{1}^{2} - T_{2}^{2}) + 4 \alpha \kappa (\beta \kappa^{2} - \gamma r_{0}^{2})(T_{1} + T_{2})}
       					   {2 \alpha^{2} \kappa^{2} \left(T_{1}^{2} + T_{2}^{2} \right) - 4 \beta \kappa^{2} (\beta \kappa^{2} - 2 \gamma r_{0}^{2}) 
		    			    + 4 \alpha \kappa (\beta \kappa^{2} - \gamma r_{0}^{2}) (T_{1} - T_{2}) + 8 \alpha^{2} \kappa^{2} T_{1} T_{2}}.}
\end{aligned}
\end{equation}

\begin{figure}[h]
\begin{center}
\begin{subfigure}{0.47\textwidth}
\includegraphics[width=\linewidth]{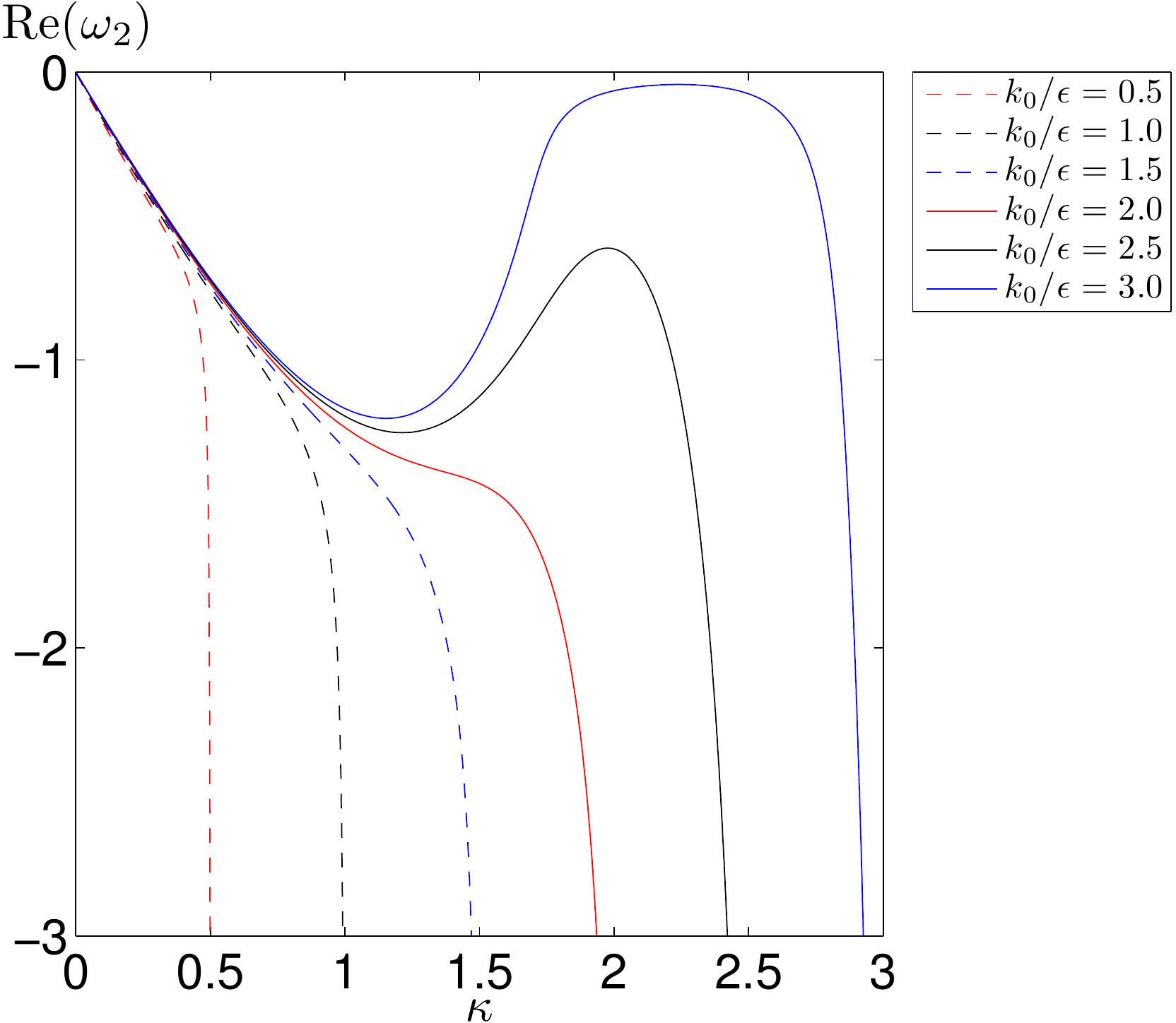}
\caption{}
\end{subfigure}
\hspace{0.5cm}
\begin{subfigure}{0.47\textwidth}
\includegraphics[width=\linewidth]{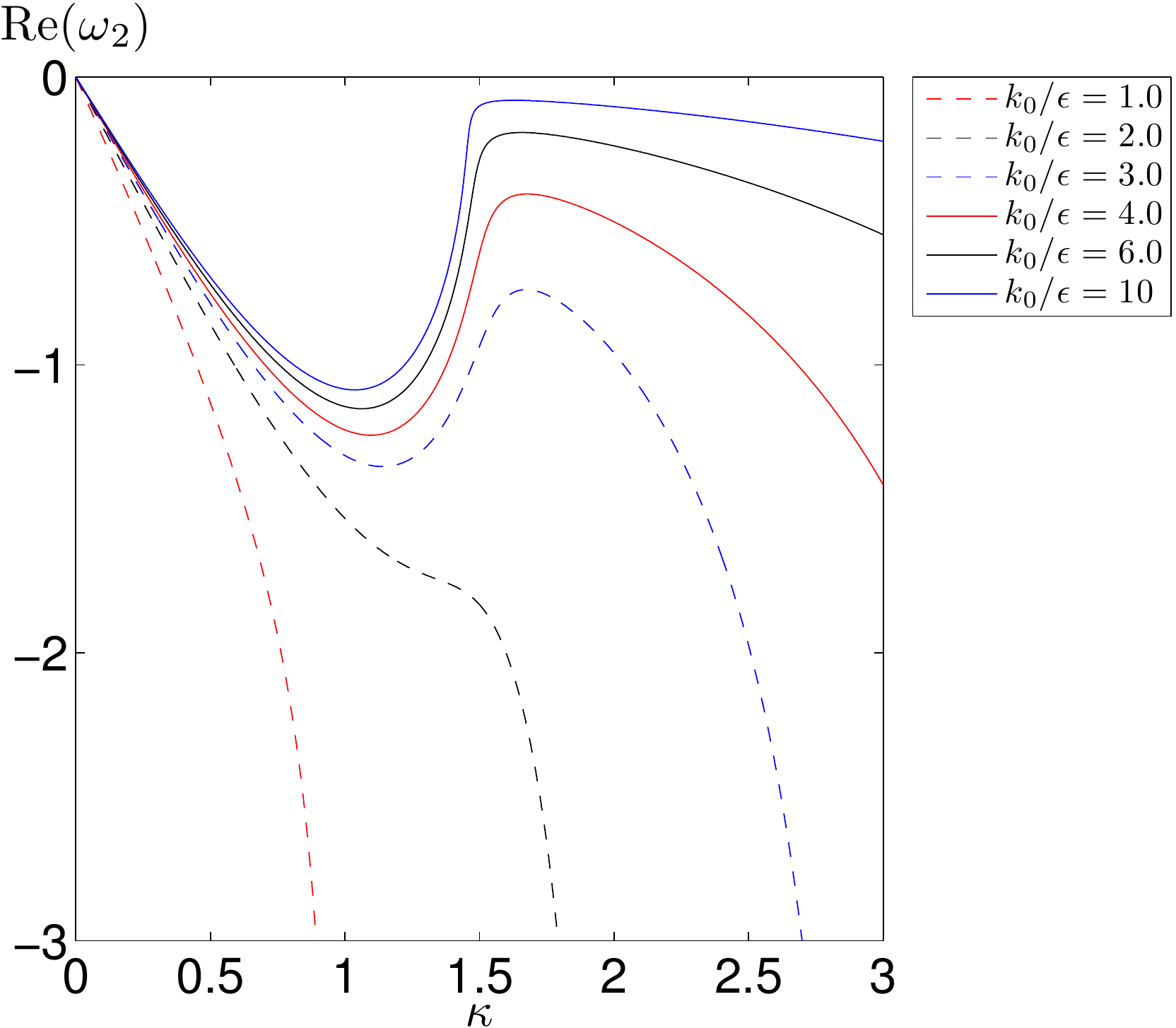}
\caption{}
\end{subfigure}
\caption{Plots of the real part of the growth rate $\omega_{2}$ as a function of the modulational wavenumber $\kappa$  for $\alpha = 1$ (a) and $\alpha = -1$ (b). The curves are given for several values of $k_{0}/\epsilon$.}
\label{neg}
\end{center}
\end{figure}

Since we are interested to investigate the stability of the plane-wave solution, we seek the values of $\kappa$ for which the real parts of $\omega_{1}$ and $\omega_{2}$ are positive, i.e. the growth rate of the instability. By taking the real part of $\omega_{1}$ and $\omega_{2}$, it turns out that for all modulation wavenumber $\kappa$ that satisfies $0 < \epsilon \kappa < k_{0}$, Re$(\omega_{1}) > 0$ and Re$(\omega_{2}) < 0$, respectively. They are given explicitly as follows:
\begin{equation}
\begin{aligned}
\textmd{Re}(\omega_{1}) &= \frac{1}{2} \left(+\sqrt{\frac{1}{2} \left[R(\kappa) + \textmd{Re}(P_{1}^{2} - 4P_{0}) \right]} - \alpha \kappa [T_{1}(\kappa) - T_{2}(\kappa)]\right), \\
\textmd{Re}(\omega_{2}) &= \frac{1}{2} \left(-\sqrt{\frac{1}{2} \left[R(\kappa) + \textmd{Re}(P_{1}^{2} - 4P_{0}) \right]} - \alpha \kappa [T_{1}(\kappa) - T_{2}(\kappa)]\right).
\end{aligned}
\end{equation}
Figure~\ref{pos} shows the real part of the growth rate $\omega_{1}$ as a function of the modulational wavenumber $\kappa$ for different values of $k_{0}/\epsilon$. Furthermore, Figure~\ref{neg} shows the real part of the growth rate $\omega_{2}$ as a function of the modulational wavenumber $\kappa$ for different values of $k_{0}/\epsilon$. Since Re$(\omega_{2}) < 0$ for $0 < \epsilon \kappa < k_{0}$, we should call it the decay rate instead of the growth rate.

\begin{figure}[h]
\begin{center}
\begin{subfigure}{0.47\textwidth}
\includegraphics[width=\linewidth]{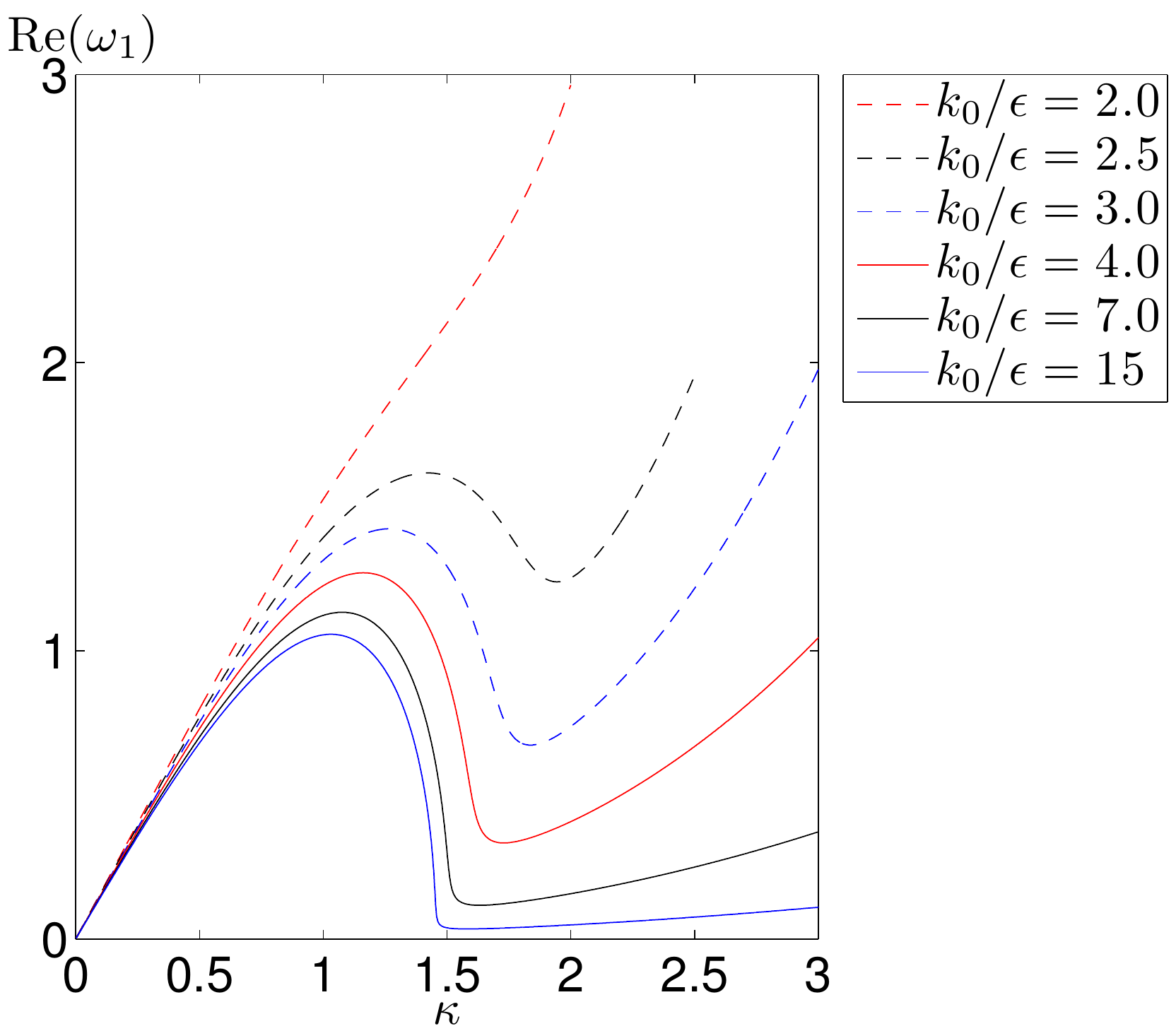}
\caption{}  
\end{subfigure}
\hspace{0.5cm}
\begin{subfigure}{0.47\textwidth}
\includegraphics[width=\linewidth]{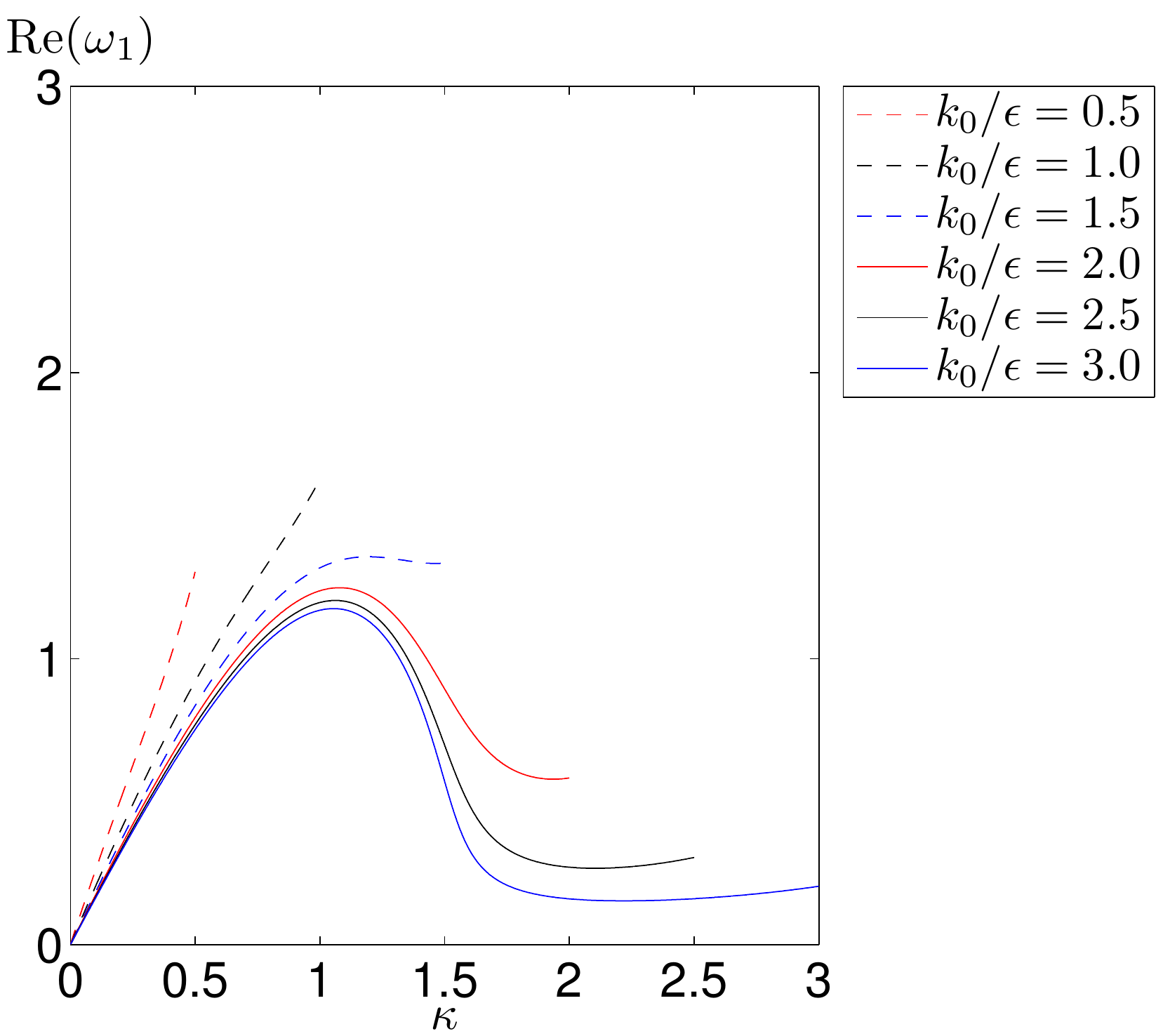}
\caption{}
\end{subfigure}
\caption{Plots of the real part of the growth rate $\omega_{1}$ as a function of the modulational wavenumber $\kappa$ for $\alpha = 1$ (a) and $\alpha = -1$ (b). The curves are given for several values of $k_{0}/\epsilon$ and under an assumption that $\kappa$ is sufficiently a small parameter.}
\label{small}
\end{center}
\end{figure}

From both figures, we observe that for the modulational wavenumber $\kappa \rightarrow k_{0}/\epsilon$, the growth rates tend to infinity. However, if we want to consider the case of narrow-banded spectra, it is desirable to restrict the modulational wavenumber $\kappa$ as a small parameter. In order to do that, we consider the Taylor expansion of $T_{1}$ and $T_{2}$ for small $\kappa$
\begin{eqnarray}
\begin{aligned}
T_{1}(\kappa) &= \sqrt{\frac{\epsilon}{2k_{0}}} \left(1 - \frac{1}{2} \frac{\epsilon}{k_{0}} \kappa + \frac{3}{8} \frac{\epsilon^{2}}{k_{0}^{2}} \kappa^2 - 
				 \dots \right), \\
T_{2}(\kappa) &= \sqrt{\frac{\epsilon}{2k_{0}}} \left(1 + \frac{1}{2} \frac{\epsilon}{k_{0}} \kappa + \frac{3}{8} \frac{\epsilon^{2}}{k_{0}^{2}} \kappa^2 + 
                 \dots \right).
\end{aligned}
\end{eqnarray}
Under this assumption, the real part of the growth rate $\omega_{1}$ as a function of the modulational wavenumber $\kappa$ for different values of $k_{0}/\epsilon$ is shown in Figure~\ref{small}. For each case of $k_{0}/\epsilon$, the growth rate curves terminate at $\kappa = k_{0}/\epsilon$. Since the growth rate is finite, this observation confirms the previous findings of Segur {\slshape et al.}~\cite{Segur05} that the dissipation, in this case, the effect of viscosity, stabilizes the modulational instability.

\section{Conclusion} \label{conclusion}

In this paper, we have derived the NLS equation modified with viscosity effect based on the KdV equation modified by viscosity from Miles~\cite{Miles76} using the method of multiple scales. Furthermore, the modulational instability of the plane-wave solution of this equation has also been investigated. It is observed that the modulational dispersion relation is a quadratic equation in $\omega$ with complex-valued coefficients. The corresponding growth rate of the NLS equation modified by viscosity shows a different pattern from the well-known Benjamin-Feir instability for the NLS equation. The stability theory is based on the assumption that we consider the case of narrow-banded spectra, that is when the modulational wavenumber is small. It is observed that for sufficiently large value modulational wavenumber $\kappa$, the instability shows an infinite pattern of the growth rate, in particular for $\kappa \rightarrow k_{0}/\epsilon$. By taking an approximate value of the growth rate for a small value of $\kappa$, we observe that the growth rate is finite and the analysis confirms the findings of Segur {\slshape et al.}~\cite{Segur05} that any type of dissipation, in this case, the effect of viscosity, stabilizes the modulational instability.

\section*{\large Acknowledgements}
The authors would like to acknowledge Professor David Benney (MIT), Professor Andy Chan (University of Nottingham Malaysia Campus), Dr. Panayotis Kevrekidis (University of Massachusetts, Amherst), Dr. Philippe Guyenne (University of Delaware, Newark), Professor Robert Conte (CMLA France and The University of Hong Kong) and Dr. Ardhasena Sopaheluwakan (Indonesian Agency for Meteorological, Climatological and Geophysics) for many fruitful discussion. This research is supported by the New Researcher Fund NRF 5035-A2RL20 from the University of Nottingham, University Park Campus, UK. The financial support from Nottingham University Business School and Faculty of Engineering, The University of Nottingham Malaysia Campus are also gratefully acknowledged.

\end{document}